\documentclass[twocolumn,aps,pra,showpacs,groupedaddress]{revtex4}
\usepackage{epsfig,amssymb,amsmath}
\def\comment#1{}\def\labell#1{\label{#1}}
\begin{document}
\title{Entropic information--disturbance tradeoff} 
\author{Lorenzo Maccone}\affiliation{QUIT - Quantum Information Theory
  Group, Dipartimento di Fisica ``A.  Volta'' Universit\`a di Pavia,
  via A.  Bassi 6, I-27100 Pavia, Italy.}

\begin{abstract}
  We show the flaws found in the customary fidelity-based definitions
  of disturbance in quantum measurements and evolutions. We introduce
  the ``entropic disturbance'' $D$ and show that it adequately
  measures the degree of disturbance, intended essentially as an
  irreversible change in the state of the system. We also find that it
  complies with an information--disturbance tradeoff, namely the
  mutual information between the eigenvalues of the initial state and
  the measurement results is less than or equal to $D$.
\end{abstract}
\pacs{03.67.-a,03.65.Yz,03.65.Ta,42.50.Lc} \maketitle 

The correct interpretation of the Heisenberg uncertainty
relations~\cite{heis,robertson} connects the uncertainty (or spread)
in the measurement results of one observable to the spread of another
observable in the initial state of the system~\cite{peres}. The true
spirit of Heisenberg's intuition~\cite{heis,mauro}, however, was that
any measurement (i.e. extraction of classical in\-for\-mation from a
quantum system) necessarily entails some kind of disturbance to the
measured system. This intuition has yet to be captured in a truly
general information--disturbance tradeoff relation. Nonetheless,
several of these relations have been put forth and cover many
conceivable situations (e.g.
see~\cite{peresfuchs,ozawanoise,mauro,bb84}).  The main problem in
deriving a general version of such a tradeoff lies in the
identification of an appropriate definition for the disturbance. 

In this paper we explore the consequences of defining the disturbance
as an irreversible change in the state of the system.  We proceed
axiomatically, by enumerating the abstract properties that a
disturbance measure should possess, and we show that the customarily
used fidelity-based disturbance measures do not satisfy them. We
introduce one that does, namely the ``entropic disturbance'' $D$. A
simple, general information--disturbance tradeoff is then derived,
namely $I\leqslant D$ ($I$ being the mutual information between the
eigenvalues of the initial state and the measurement results). Its
interpretation is straightforward: Every time an amount $I$ of
information is obtained from the measurement apparatus, a disturbance
$D$ at least as big is introduced on the system, but, obviously, the
system can be disturbed by a process that returns little information.
We also derive the equality conditions $I=D$ showing that the
$I\leqslant D$ bound is tight, and proving (as is to be
expected~\cite{rudolph}) that von Neumann-type measurements are among
the least disturbing amid all measurements that retrieve the same
information. Finally, we give a state--independent tradeoff by
averaging over all possible input states.  This provides a measure of
the global disturbance introduced by the apparatus or by a generic
evolution.

\section{Measurement}
A measurement is by definition an operation that acts on a system and
returns some classical information, i.e. a label ``$l_k$'' identified
by an index $k$. This operation typically, but not always, changes the
state of the measured system ({\it wave-function collapse} mechanism).
When measuring classical objects, the state change is only due to the
change in the information we have on the system: The uncertainty in
the state is usually reduced. When measuring quantum objects, the
state change can have a dynamical nature that perturbs the system.
The rigorous description of these mechanisms results from the Kraus
decomposition of the measurement apparatus~\cite{kraus,ozawanoise},
which identifies the completely positive (CP) map of its evolution.
When the system is initially in a state $\rho$, the $k$th measurement
outcome occurs with a probability $p_k=$Tr$[\Pi_k\rho]$, where
$\{\Pi_k\}$ is the apparatus POVM (Positive Operator-Valued Measure),
a set of positive operators normalized so that $\sum_k\Pi_k=\openone$.
After the outcome $l_k$ is obtained, the state is changed to
\begin{eqnarray}
\rho'(k)=\sum_{i\in I_k}K_i\:\rho\:{K_i}^\dag/p_k
\;\labell{statered},
\end{eqnarray}
where $I_k$ is a set of indices $i$ and $K_i$ are the apparatus Kraus
operators. The POVM in terms of these is given by $\Pi_k=\sum_{i\in
  I_k}{K_i}^\dag K_i$.  On the basis of the above description of the
measurement, we can define the following information and disturbance
measures.

{\em Information:} The outcome $l_k$ provides the experimenter with
some information $I$. We are, obviously, interested in the case in
which $l_k$ provides some information on the measured system, and it
is not just independently generated by the apparatus. Thus, a good
measure for $I$ is the mutual information between the measurement
results and some property of the system state. A significant
(basis-independent) property is the spectrum of the state $\rho$, i.e.
the probability distribution of its eigenvalues $\lambda_j$. Of
course, the ensemble composed by the eigenvectors of $\rho$ weighted
by the corresponding eigenvalues is not the only one that originates
from the state $\rho$. However, it is easy to see that it is the one
that allows to recover the maximal accessible information (it
saturates the Holevo bound)~\cite{wootters}.  It is thus the
appropriate choice since we ultimately want to bound the information
from above.  We then use
\begin{eqnarray}
I\equiv I(\lambda_j,p_k)\labell{defi}\;,
\end{eqnarray}
the mutual information between the label $j$ identifying the
eigenvalue of $\rho$ and the label $k$ identifying the measurement
result: $I$ is the number of bits that the experimenter gains from the
result $l_k$ on which eigenvector of $\rho$ the system was in (before
the measurement).  \comment{ATTENZIONE!! $I$ non \`e particolarm
  rilevante, perche' la ``vera'' $\rho$ (quella di Alice) avra' avett
  diversi!  \`E comunque un lower bound per l'info che posso ottenere
  su quale\`e il ``vero'' stato (bound achievable se per caso la vera
  $\rho$ ha gli stessi avett).}  It is maximal when all eigenvalues
are equal (the experimenter has no prior info on the state), and it is
null when the state is pure (the experimenter already has total
knowledge of the state, and cannot gain any more information on it).
[Note that $\rho$ here refers to the state from the experimenter's
point of view: It reflects his prior knowledge of the system state,
i.e. that each eigenvector has a prior probability $\lambda_j$.  The
`true' state (i.e. from the point of view of who is preparing the
system) will be in general purer. The knowledge that can be acquired
by the experimenter is upper bounded by the difference in entropy
between these two representations of the state.\comment{Cioe' l'info
  che Bob puo' ottenere sullo stato di Alice \`e $I_B\leqslant
  S(\rho)-S(\rho_A)$.\`E possibile calcolare l'info che B ottiene dati
  $\rho$ e $K_i$? NO! Avro' bisogno anche di $\rho_A$: fissato $K_i$,
  al variare di $\rho_A$ ottengo info ben diverse!}]  If the Hilbert
space of the system has finite dimension $d$, we can normalize $I$ by
dividing it with its maximum value $\log_2d$, so to have $0\leqslant
I\leqslant 1$.

{\em Disturbance:} A disturbance is an irreversible change in the
state of the system, caused by a CP-map evolution (such as the
dynamical disturbance caused by quantum correlations that leak out to
the environment and are lost).  Thus, any quantity $D$ that measures
disturbance should satisfy the following requirements, inspired by
Ref.~\cite{mauro}:
\begin{itemize}
\item[{\em i)}]$D$ should be a function only of the input state
  $\rho$ and of the apparatus, identified through its Kraus operators
  $\{K_i\}$, i.e.  $D=D(\rho,\{K_i\})$. 
\item[{\em ii)}]$D$ should be null if and only if the transformation
  $\{K_i\}$ is invertible on $\rho$~\footnote{Note that by
    ``invertible'' we mean that there exists a transformation that,
    acting on the final state of the system only, is able to recover
    the initial state even if the system was initially entangled with
    some other system. This means that if $|\Psi\rangle$ is a
    purification of $\rho$, there exists an ``inversion'' CP-map
    identified by the Kraus operators $\{I_j\}$ such that
    $|\Psi\rangle\langle\Psi|=\sum_{ij}(\openone\otimes
    I_jK_i)|\Psi\rangle\langle\Psi|(\openone\otimes
    {K_i}^\dag{I_j}^\dag)$, where the identity $\openone$ acts on the
    purification space. }. In this case the state change can be
  undone, and such transformation is not disturbing the system.
\item[{\em iii)}]Once the state has been disturbed, it should not be
  possible to decrease $D$ with any successive transformation.  This
  means that $D$ should be monotonically non-decreasing for successive
  applications of CP-maps~\cite{rudolph} (i.e. it should satisfy a
  sort of pipeline inequality). This requirement captures the notion
  that a disturbance should be irreversible\comment{Togli il seguente
    (e la citaz) se serve spazio}, and is connected with the concept
  of ``cleanness''~\cite{werner}.  
\item[{\em iv)}]$D$ should be continuous: Maps and input states which
  do not differ too much should give similar values of $D$.
\end{itemize}

The above requirements, which define the disturbance axiomatically,
have nothing to do with the information the measurement provides. As
such, there is no obvious {\it a priori} reason why an
information--disturbance tradeoff should hold.

Definitions of disturbance are customarily based on the fidelity or
the Bures distance~\cite{chuang} between input and output states. Even
though valid information--disturbance relations can be
found~\cite{peresfuchs}, these definitions do not seem to
appropriately gauge the disturbance, intended as an irreversible
evolution. In fact, even though a unitary transformation is perfectly
reversible, it can rotate a state to an orthogonal configuration,
generating the maximum possible fidelity-based disturbance. These
quantities do not satisfy the requirements {\em ii)} and~{\em iii)}.
Analogous considerations apply also if we use the entanglement
fidelity~\cite{efidel} in place of the fidelity~\footnote{One could
  try enforcing requirements {\em ii)} and~{\em iii)} by maximizing
  the fidelity over all possible unitary operators, defining a
  disturbance of the form $\bar D=1-\max_{\bar U} F(\rho,\bar
  U\:\rho'\:{\bar U}^\dag)$, where $F$ is the fidelity, $\rho$ and
  $\rho'$ are the input and output states, and the maximization runs
  over all unitaries $\bar U$. Also this definition is inadequate,
  since $\bar D$ is null if $\rho$ and $\rho'$ have the same
  eigenvalues, which does not necessarily entail that the
  transformation is invertible, i.e.  requirement {\em ii)} still does
  not hold.}.

A definition of disturbance $D$ that satisfies all the above
requirements can be found by recalling that a CP-map $\cal Q$ is
invertible if and only if~\cite{cnes} the map's coherent information
$I_c(\rho,{\cal Q})$ is equal to the von Neumann entropy
$S(\rho)=-$Tr$[\rho\log_2\rho]$ of the input state $\rho$.  The
coherent information~\cite{seth,cnes} is defined as $I_c\equiv
S\left({\cal Q}(\rho)\right)-S\left(({\cal
    Q}\otimes\openone)(|\Psi\rangle\langle\Psi|)\right)$, where
$|\Psi\rangle$ is a purification of $\rho$ and the map ${\cal
  Q}\otimes\openone$ acts with $\cal Q$ on the system space and with
the identity $\openone$ on the purification space. The quantity $I_c$
is non-increasing for application of CP-maps (data-processing
inequality)~\cite{cnes}. Namely, for any two maps $\cal Q$ and $\cal
Q'$, we have $I_c(\rho,{\cal Q})\geqslant I_c(\rho,{\cal Q'}\circ{\cal
  Q})$, where $\circ$ denotes composition of maps. Thus, a disturbance
measure that satisfies requirements~{\em ii)-iii)} must be a function
$f$ of $S(\rho)-I_c$, with $f$ non-decreasing and null when its
argument is: $f(0)=0$.  We then define \begin{eqnarray} D&\equiv&
  S(\rho)-I_c\labell{defdist}\\&=&S(\rho)- S\left({\cal
      Q}(\rho)\right)+S\left(({\cal
      Q}\otimes\openone)(|\Psi\rangle\langle\Psi|)\right) \nonumber,
\end{eqnarray}
which, in addition to~{\em ii)-iii)}, also satisfies requirements~{\em
  i)} and~{\em iv)} since it is continuous (see the Appendix).
Analogously to $I$, also $D$ can be normalized in $d$-dimensional
Hilbert spaces by dividing it by $\log_2d$, so that $0\leqslant
D\leqslant 2$.

From the postulates of quantum mechanics it follows that the system
state $\rho$ describes the information the experimenter possesses on
the system. Hence, there are two mechanisms that lead to a state
change: the system dynamics and the acquisition of new information.
Heisenberg, in his uncertainty principle, was considering the former
mechanism only. To exemplify the latter, suppose I acquire a qubit in
an unknown state.  Initially, I will assign to it the state
$\openone/2$, but as soon as the preparer tells me that the qubit was
in the state $\rho_p$, from my point of view it undergoes a state
change (even though I may have not interacted with it) described by
the map $\rho_p={\cal C}[\openone/2]$.  We can call this a ``purely
informational'' state change. Since both the dynamical and the
informational state changes are described by CP-maps, they both fall
in the general framework described above. [In this sense it may be
interpreted as a generalization of Heisenberg's intuition]. Is it
possible to weight the contribution of these two mechanisms in each
measurement apparatus?  Yes: Since the set of CP-maps is a convex set,
the apparatus CP-map $\cal Q$ (identified by the Kraus operators
$\{K_i\}$) can always be written as a convex combination of the purely
informational map $\cal C$ and of a ``dynamical'' map $\cal T$ as
\begin{eqnarray}
{\cal Q}=\xi{\cal C}+(1-\xi){\cal T}\mbox{ with }\xi\in[0,1]
\labell{sdas}\;.
\end{eqnarray}
 The Kraus
operators of the map $\cal C$ are
$A_{jk}\equiv\sqrt{\mu_j}|v_j\rangle\langle v_k|$, where $\mu_j$ and
$|v_j\rangle$ are the eigenvalues and eigenvectors of the `true' (i.e.
from the point of view of the preparer) state $\rho_p$: The action of
$\cal C$ must not change the true state, ${\cal C}[\rho_p]=\rho_p$.
[Note that in the case of a degenerate state $\rho_p$, the map $\cal
C$ might not be univocally defined]. The POVM of a purely
informational measurement is then $\Pi_k=\sum_j{A_{jk}}^\dag
A_{jk}=|v_k\rangle\langle v_k|$ (the projectors on the eigenspaces of
$\rho_p$). \comment{Sembrerebbe che tutte le informational maps siano
  invertibili! NO! La mappa $\cal C$ non porta il mio stato nel vero
  stato $\rho$ ($\rho={\cal C}[\rho]$), ma porta TUTTI gli stati in
  $\rho$. Questo non\`e invertibile.} In this respect, the {\em truly
  quantum} contribution to the disturbance in a measurement is related
to the map $\cal T$: it is present in those measurement apparatuses
with $\xi<1$.  \comment{ATTENZIONE! Forse questa property non
  identifica univocally la mappa $\cal C$ e quindi questo non
  determina univocally la $\xi$!!!!-->S\`I}

\begin{figure}[hbt]
\begin{center}
\epsfxsize=.85\hsize\leavevmode\epsffile{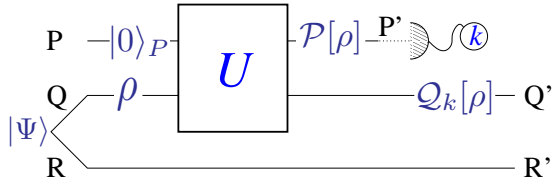}
\end{center}
\caption{Measurement apparatus described through the indirect
  measurement model. Any apparatus can be described~\cite{ozawanoise}
  by coupling the system to be measured (in the space Q) with a probe
  P through a unitary operator $U$. The probe is then projectively
  measured at the output P$'$.  Notation employed below:
  $|0\rangle_P=$~initial state of the apparatus in P; $\rho=$~initial
  state of the system in Q; ${\cal P}[\rho]=$~state of the probe in
  P$'$, before the final projective measurement; ${\cal
    Q}_k[\rho]=$~state of the system in Q$'$, after the interaction
  with the apparatus and after the probe measurement with result $l_k$
  (i.e.  after the wave-function collapse). The reference R is
  introduced for purification purposes: It is defined so that the
  joint initial state $|\Psi\rangle$ of system and reference, QR, is
  pure.}  \labell{f:imm}\end{figure}

\section{Information--disturbance tradeoff} 
We now prove the information--disturbance tradeoff \comment{la seguente
  equaz puo' essere messa in linea, eliminando la citazione sotto}
\begin{eqnarray}
I(\lambda_j,p_k)\leqslant D(\rho,\{K_i\})
\;\labell{infodis}.
\end{eqnarray}
Any measurement apparatus can be decomposed into a unitary evolution
$U$ (the Stinespring dilation of the apparatus) followed by a von
Neumann projective measurement on a probe P, the so-called {\em
  indirect measurement model}~\cite{ozawanoise}, see Fig.~\ref{f:imm}.
The unitary $U$ couples Q with the probe P in the apparatus yielding
Q$'$ and P$'$.  The joint evolution of probe and system PQ in the
apparatus can be seen as composed by two complementary quantum
channels: A channel Q$\rightarrow$P$'$ that describes the transfer of
information from the system to the state of the probe in P$'$ which is
then measured yielding the measurement result $l_k$, and a channel
Q$\rightarrow$Q$'$ that evolves the system before the measurement into
the system after the measurement conditioned on its result $l_k$.
These two channels are respectively described by the CP-maps ${\cal
  P}[\rho]\equiv$Tr$_Q[U(\rho\otimes|0\rangle_P\langle 0|)U^\dag]$ and
${\cal Q}_k[\rho]\equiv$Tr$_P\big[U(\rho\otimes|0\rangle_P\langle
0|)U^\dag(\openone_Q\otimes|k\rangle_P\langle k|)\big]/p_k$, where
$|0\rangle_P$ is the initial pure state of the probe, $|k\rangle_P$ is
the basis representing the projective measurement on the probe, and
$p_k$ is the probability of the $k$th result. Since the final state of
the system in Q$'$ is given by $\rho'(k)={\cal Q}_k[\rho]$ with
probability $p_k$, it can be written as
$\sum_kp_k\rho'(k)=$Tr$_P[U(\rho\otimes|0\rangle_P\langle
0|)U^\dag]\equiv{\cal Q}[\rho]$, where the Kraus operators of the map
$\cal Q$ can be immediately obtained from the ones of the maps ${\cal
  Q}_k$.  The map $\cal Q$ describes the unitary coupling of system
and probe in the apparatus and the successive trace on the probe
space, which yields the unconditioned output state.

The system's initial state $\rho$, expanded on its eigenvectors
$|j\rangle$ is given by $\rho=\sum_j\lambda_j|j\rangle\langle j|$.  We
use the Holevo-Schumacher-Westmoreland theorem~\cite{hsw} with an
alphabet composed by $|j\rangle\langle j|$ with probability
$\lambda_j$ flowing through a channel described by $\cal P$. Such
theorem implies that the mutual information $I(\lambda_j,p_k)$ between
the variable $j$ and the measurement results $k$ (whatever measurement
strategy is employed) is upper bounded as
\begin{eqnarray}
I(\lambda_j,p_k)\leqslant S({\cal P}[\rho])-\sum_j\lambda_j
S\big({\cal 
  P}\big[|j\rangle\langle j|\big]\big)
\;\labell{hsw}.
\end{eqnarray}
The system in space Q can be purified by adding an auxiliary reference
space R, so that the system in QR is initially in a pure state
$|\Psi\rangle$. The entropy $S(P')\equiv S({\cal P}[\rho])$ of the
probe just before the final von Neumann measurement is then equal to
the entropy $S($Q$'$R$')$ of the joint state in Q$'$R$'$ of the system
and the reference after the interaction $U$. In fact, the initial
state in PQR is pure and it is evolved into P$'$Q$'$R$'$ by a unitary
evolution.  Thus,
\begin{eqnarray}
S({\cal P}[\rho])=S(\mbox{Q}'\mbox{R}')=S_e(\rho,{\cal Q})
\;\labell{ppp1},
\end{eqnarray}
where $S_e(\rho,{\cal Q})$ is the exchange entropy~\cite{seth,efidel}
of the map $\cal Q$. It is defined as the entropy of the joint
Q$'$R$'$ output state of system and purification-reference, i.e.
$S_e(\rho,{\cal Q})\equiv S\big(({\cal
    Q}\otimes\openone_R)\big[|\Psi\rangle\langle\Psi|\big]\big)$.  Moreover,
the entropy in Q$'$ satisfies
\begin{eqnarray}
&&S(\mbox{Q}')\equiv S({\cal Q}[\rho])
\labell{ppp2}\\\nonumber&&=
S\Big(\sum_j\lambda_j{\cal
  Q}\big[|j\rangle\langle j|\big]\Big)\leqslant
\sum_j\lambda_j S\big({\cal
  Q}\big[|j\rangle\langle j|\big]\big)+H(\lambda_j)
,
\end{eqnarray}
where $H(\lambda_j)=S(\rho)$ is the Shannon entropy of the probability
distribution $\lambda_j$, and where the inequality
$S(\sum_xp_x\varrho_x)\leqslant\sum_xp_xS(\varrho_x)+H(p_x)$ (valid
for all probabilities $p_x$ and states $\rho_x$) has been
used~\cite{chuang}. Notice that if the system in Q is initially in a
pure state $|j\rangle$, the entropy of the output of the two channels
$\cal P$ and $\cal Q$ coincides since the entropy of the joint system
PQ is initially null.  Hence, $S\big({\cal Q}\big[|j\rangle\langle
j|\big]\big)=S\big({\cal P}\big[|j\rangle\langle j|\big]\big)$, so
that Eq.~(\ref{ppp2}) implies
\begin{eqnarray}
\sum_j\lambda_jS\big({\cal P}\big[|j\rangle\langle
j|\big]\big)\geqslant S({\cal 
  Q}[\rho])-S(\rho)
\;\labell{ppp3}.
\end{eqnarray}
Joining Eqs.~(\ref{hsw}),~(\ref{ppp1}) and~(\ref{ppp3}), we find
$I\leqslant S_e(\rho,{\cal Q})-S({\cal
  Q}[\rho])+S(\rho)=S(\rho)-I_c(\rho,{\cal Q}) =D$, thus proving
Eq.~(\ref{infodis}).  Notice that such proof works also in the case in
which the input and output Hilbert spaces Q and Q$'$ do not coincide,
and in the case of infinite dimensional Hilbert spaces~\cite{yuen}.

\comment{Tutto il seguente forse si puo' eliminare: rischia solo di
  fare confusione e far sembrare $D$ una misura di info e non di
  disturbance! (se eliminato, va tolta anche la referenza qcapacity).}
There is a simple, not-very-rigorous, intuition behind the preceding
proof.  The total quantum information of the initial state $\rho$ can
be quantified by $N\simeq S(\rho)$ qubits. The unitary evolution $U$
transfers $n$ of them to the probe space P, where the projective
measurement can return a number of bits $b\leqslant n$, due to the
Holevo bound~\cite{chuang}.  The remaining $N-n$ qubits constitute an
upper bound to the quantum capacity to transfer the quantum
information in the initial state through the channel
Q$\rightarrow$Q$'$ consisting of the measurement apparatus.  The
quantum capacity is measured by the coherent
information~\cite{seth,qcapacity}, so that $I_c\leqslant N-n$. Thus,
\begin{eqnarray}
I\simeq b\leqslant n=N-(N-n)\leqslant N-I_c\simeq D\;.
\labell{ssss}\;
\end{eqnarray}

We now deduce the equality conditions for the information--disturbance
bound $I\leqslant D$, showing that it is achievable. The equality in
the Holevo-Schumacher-Westmoreland relation of Eq.~(\ref{hsw}) is
achieved if the alphabet states ${\cal P}\big[|j\rangle\langle
j|\big]$ commute~\cite{comm}. Moreover, the equality in the relation
$S(\sum_xp_x\varrho_x)\leqslant\sum_xp_xS(\varrho_x)+H(p_x)$, which
was employed in Eq.~(\ref{ppp2}), is achieved if and only if the
states $\rho_x$ have support on orthogonal subspaces~\cite{chuang}
(which implies that they commute). Thus, we have equality $I=D$ if and
only if the channel $\cal P$ maps different eigenvectors $|j\rangle$
of the initial state $\rho$ into orthogonal subspaces. A typical
example is a projective measurement whose Kraus operators are
projectors on the basis $|j\rangle$. It is a purely informational
measurement, where the only uncertainty derives from classical
probability. [Notice that, in a $d$-dimensional Hilbert space, the
converse also holds for measurements with $d$ outcomes: If $I=D$ and
the measurement POVM has $d$ elements, then the measurement is a von
Neumann-type projection, i.e. its Kraus operators are of the form
$A_j=\tilde U|a_j\rangle\langle a_j|$ where $\tilde U$ is a fixed
unitary and $|a_j\rangle$ is a basis. The proof of this assertion
follows immediately from the fact that the measurement must map a
basis $|j\rangle$ of $d$ elements into $d$ orthogonal subspaces,
which, in a $d$-dimensional Hilbert space, must then be
one-dimensional].

\section{State--independent tradeoff} 
The definitions we used for information $I(\lambda_j,p_k)$ and
disturbance $D(\rho,\{K_i\})$ are explicitly dependent both on the
input state $\rho$ and on the apparatus. We can forgo the state
dependence by averaging on all possible input states with equal
weights (for symmetry reasons), i.e.  by using a state
$\rho=\openone/d$ in a $d$-dimensional Hilbert space.  [In
infinite-dimensional spaces, additional requirements are also
necessary, such as using states with upper-bounded energy.] Thus, we
can define a state--independent information as $\widetilde I(p_k)\equiv
I(\openone/d,p_k)$ and a state--in\-de\-pen\-dent disturbance as
$\widetilde D(\{K_i\})\equiv
D(\openone/d,\{K_i\})=\log_2d-I_c(\openone/d,{\cal Q})$.  The quantity
$\widetilde I/\log_2d$ measures the percentage of the maximum
retrievable information that is achieved by the apparatus.  The
quantity $\widetilde D$ measures the disturbance the apparatus causes
to a completely unknown state. Notice that $\widetilde D=0$ if and
only if the apparatus acts on the state with a unitary transformation
(i.e. it yields no information on the state).  In fact,
$S(\rho)=I_c(\rho,{\cal Q})$ if and only if the map $\cal Q$ is
invertible on all the pure states in the support of
$\rho$~\cite{cnes}, which for $\rho=\openone/d$ implies that the map
is unitary.  \comment{No: il seguente e' sbagliato: Cpe, una misura
  proiettiva mi sa che ha ancora $\widetilde I=\widetilde D$...
  Moreover, from the equality condition we derived above, it follows
  that the state--independent tradeoff $\widetilde I\leqslant\widetilde
  D$ is tight only in this case, i.e. when $\widetilde I=\widetilde
  D=0$. }

\section{Conclusions} 
We have introduced a new measure $D(\rho,\{K_i\})$ of the disturbance
(intended as an irreversible state change) that a map with Kraus
operators $\{K_i\}$ induces on a system in a state $\rho$.  We have
derived an information--disturbance tradeoff for such a quantity in
the form $I\leqslant D$, where $I$ is the classical info the map
$\{K_i\}$ returns on the state $\rho$ to the experimenter. The
equality conditions for this bound have been also derived. Moreover, a
state--independent tradeoff $\widetilde I\leqslant\widetilde D$ was
obtained, which bounds the percentage of the maximum achievable
information $\widetilde I$ with the disturbance $\widetilde D$ caused
to a completely unknown input state.

Even though we found a valid information--disturbance tradeoff, we
have to conclude that the disturbance definition $D$ we used here may
not be the appropriate definition to capture the true spirit of
Heisenberg's intuition (the ``uncertainty principle''). In fact, a
``classical'' measurement (or a purely informational state change)
where one gains information on which, out of a set of orthogonal
configurations, our state is in, will perturb the state in an
irreversible manner even though there is no dynamical interaction on
the system. Heisenberg, on the other hand, analyzed only the
irreversible state changes induced by dynamical actions (measurements)
on the system.  In this sense, the information--disturbance relation
derived here might be considered as an extension of Heisenberg's
intuition on the disturbance a measurement induces.

\appendix\section{Continuity of $D$} 
Here we prove that the entropic disturbance $D$ is continuous. More
rigorously, we prove the following two statements: {\em
  i)}~$\rho\to\rho'$, i.e. $T(\rho,\rho')\to 0$, implies $D(\rho,{\cal
  Q})\to D(\rho',{\cal Q})$, where the trace distance $T$ is defined
as $T(\rho,\rho')\equiv$Tr$\big[|\rho-\rho'|\big]/2$; {\em ii)}~
${\cal Q}\to{\cal Q}'$, i.e.  $T({\cal Q}[\rho],{\cal Q}'[\rho])\to
0$, implies $D(\rho,{\cal Q})\to D(\rho,{\cal Q}')$.

Proof of {\em i)}: Start from 
\begin{eqnarray}
&&D(\rho,{\cal Q})-D(\rho',{\cal
  Q})=[S(\rho)-S(\rho')]\labell{st}\\\nonumber
&&-[S({\cal Q}(\rho))-S({\cal
  Q}(\rho'))]+[S_e(\rho,{\cal Q})-S_e(\rho',{\cal Q})]
\;.
\end{eqnarray}
The first bracket in Eq.~(\ref{st}) tends to zero for $\rho\to\rho'$
thanks to the continuity of the entropy. It derives from Fannes'
inequality~\cite{fannes}, according to which
$|S(\rho)-S(\rho')|\leqslant h(T(\rho,\rho'))$ with the function
$h(x)\to 0$ for $x\to 0$.  The second bracket in Eq.~(\ref{st})
analogously goes to zero since it is bounded by the first: The
contractivity of CP-maps~\cite{maryb} implies that   $T({\cal
Q}(\rho),{\cal Q}(\rho'))\leqslant T(\rho,\rho')$.  To show that also
the last bracket tends to zero, recall that the exchange entropy can
be written as $S_e(\rho,{\cal Q})=S(W)$ with the matrix $W$ defined by
$W_{ij}=$Tr$[K_i\rho K_j^\dag]$, $K_i$ being the Kraus operators of
$\cal Q$~\cite{efidel}.  Thus, Fannes' inequality implies
$|S_e(\rho,{\cal Q})-S_e(\rho',{\cal Q})|\leqslant h(T(W,W'))$, and
for $\rho\to\rho'$, we have $W\to W'$.  In fact, since $W$ is
Hermitian,
\begin{eqnarray}
|W_{ij}-W'_{ij}|^2&=&\mbox{Tr}[A_j^\dag
A_i(\rho-\rho')]\mbox{Tr}[A_i^\dag
A_j(\rho-\rho')]\nonumber\\&\leqslant&
\mbox{Tr}[A_j^\dag A_iA_i^\dag 
A_j]\mbox{Tr}[(\rho-\rho')^2]
\;\labell{fin},
\end{eqnarray}
where we used the Schwarz inequality for the Hilbert-Schmidt scalar
product of operators: $\langle A|B\rangle\equiv$Tr$[A^\dag B]$.

Proof of {\em ii)}: it follows immediately from the continuity of the
entropy, i.e. from the Fannes inequality~\cite{fannes}.

I thank G. Chiribella, G. M. D'Ariano and V. Giovannetti for
stimulating hints and discussions. Financial support comes from MIUR
through FIRB (bando 2001) and PRIN 2005.


\begin{references}
\bibitem{heis} W. Heisenberg, Zeitsch. Phys. {\bf 43}, 172 (1927),
  English translation: J. A.  Wheeler and H. Zurek, {\em Quantum
    Theory and Measurement} (Princeton Univ.  Press, 1983), pg.  62.
\bibitem{robertson}H. P. Robertson, Phys. Rev. {\bf 34}, 163 (1929).
\bibitem{peres}A. Peres {\em Quantum Theory: Concepts and Methods}
  (Kluwer ac. publ., Dordrecht, 1993).
\bibitem{mauro} G. M. D'Ariano, Fortschr. Phys. {\bf 51}, 318 (2003).
\bibitem{peresfuchs}C. A. Fuchs and A. Peres, Phys. Rev. A {\bf 53},
  2038 (1996); C. A. Fuchs, Fortschr. Phys. {\bf 46}, 535 (1998); H.
  Barnum,  quant-ph/0205155 (2002); L. Maccone, Phys. Rev. A
  {\bf 73}, 042307 (2006).
\bibitem{ozawanoise}M. Ozawa, Ann. Phys. {\bf 311}, 350 (2004).
\bibitem{bb84} C. H. Bennett, G. Brassard, and N. D. Mermin, Phys.
  Rev. Lett. {\bf 68}, 557 (1992); M. Ban, J. Phys. A: Math. Gen. {\bf
    32}, 6527 (1999); K. Banaszek, Phys. Rev. Lett. {\bf 86}, 1366
  (2001); P.  Arrighi, Int. J. Quantum Inf.  {\bf 2}, 341 (2004); F.
  Sciarrino, M. Ricci, F. De Martini, R.  Filip, and L.  Mi\v sta, Jr.
  Phys. Rev. Lett. {\bf 96}, 020408 (2006); M.  Christandl and A.
  Winter, quant-ph/0501090 (2005).
\bibitem{rudolph} S. D. Bartlett, T. Rudolph, R.  W. Spekkens, P. S.
  Turner, New J. Phys. {\bf 8}, 58 (2006).
\bibitem{kraus} K. Kraus {\em States Effects and Operations}
  (Springer-Verlag, Berlin, 1983); K. Kraus, Phys. Rev. D {\bf 35},
  3070 (1987).
\bibitem{wootters}R. Jozsa, D. Robb, and W. K. Wootters, Phys. Rev. A
  {\bf 49}, 668 (1994).
\bibitem{werner} F. Buscemi, G. M. D'Ariano, M. Keyl, P. Perinotti, and
  R. F. Werner, J. of Math.  Phys. {\bf 46}, 082109 (2005).
\bibitem{chuang} M. A. Nielsen and I. L. Chuang, {\em Quantum
    Computation and Quantum Information} (Cambridge Univ. Press,
  Cambridge, 2000).
\bibitem{efidel}B. Schumacher, Phys. Rev. A {\bf 54}, 2614 (1996). 
\bibitem{cnes}B. Schumacher and M. A. Nielsen, Phys. Rev. A {\bf 54},
  2629 (1996).
\bibitem{seth}S. Lloyd, Phys. Rev. A {\bf 55}, 1613 (1997).
\bibitem{hsw} A. S. Holevo, IEEE Trans. Inf. Theory {\bf 44}, 269
  (1998); B. Schumacher and M. D. Westmoreland, Phys. Rev. A {\bf 56},
  131 (1997).
\bibitem{yuen} H. P. Yuen, M. Ozawa, Phys. Rev. Lett. {\bf 70}, 363
   (1992).
\bibitem{qcapacity} H.  Barnum, M. A. Nielsen, and B. Schumacher,
   Phys. Rev. A {\bf 57}, 4153 (1998); I. Devetak, IEEE Trans. Inform.
   Theory {\bf 51}, 44 (2005).
\bibitem{comm}M. J. W. Hall, Phys. Rev. A {\bf 55}, 100 (1997).
\bibitem{fannes}M. Fannes, Commun. Math. Phys. {\bf 31}, 291 (1973).
\bibitem{maryb}M. B. Ruskai, Rev. Math. Phys. {\bf 6}, 1147 (1994).
\end{references}
\end{document}